# POSITRON ACCELERATION IN THUNDERSTORMS

Chilingarian, B. Sargsyan

A I Alikhanyan National Lab (Yerevan Physics Institute)

Alikhanyan Brothers 2, Yerevan, Armenia AM0036

## Abstract

Thunderstorm Ground Enhancements (TGEs) are known manifestations of relativistic runaway electron avalanches (RREAs) developing inside thunderclouds. However, the role of positrons in TGEs and their relationship to thundercloud charge structure remain poorly understood. We report time-resolved observations of intense positron fluxes detected at the Aragats Observatory on 16-17 May 2026 during strong thunderstorms. Two of the three events were characterized by a positive near-surface electric field (NSEF), graupel precipitation, a low cloud base, and low-to-moderate enhancement of gamma-ray/electron fluxes measured by SEVAN and STAND3 detectors. The third event is a classical electron TGE considered for comparative purposes. All events show moderate to strong enhancement of the 511-keV annihilation line, along with enhanced radon progeny gamma-ray lines. We observe a temporal separation between the electron/gamma-ray TGE peak and the positron flux maximum. To explain these observations, we introduce a dual-dipole electrodynamic model. The large-scale electron dipole comprises the main negative thundercloud layer and its broad positive mirror charge induced at the Earth's surface, producing ordinary TGEs. Simultaneously, a localized positron dipole forms from the LPCR, with its negative mirror charge directly beneath the LPCR footprint. This lower dipole accelerates positrons downward while decelerating electrons entering the same region. These results establish positron TGEs as a new subclass of atmospheric high-energy phenomena and provide direct evidence of localized positron acceleration in the lower atmosphere.

## Plain Language Summary

Thunderstorms are natural particle accelerators that produce bursts of high-energy radiation called Thunderstorm Ground Enhancements (TGEs). Previous studies mainly focused on electrons and gamma rays generated inside thunderclouds. In this work, we show that thunderstorms can also locally accelerate positrons — the antimatter counterparts of electrons. During strong thunderstorms on 16–17 May 2026, detectors at the Aragats high-altitude research station in Armenia measured an unusual enhancement of 511-keV gamma rays, produced when positrons annihilate with electrons. The strongest positron signal appeared after the ordinary thunderstorm particle burst had peaked, indicating that positrons are governed by a different part of the thundercloud's electric structure. We propose that mature lower positive charge regions (LPCRs), formed near the cloud base alongside graupel precipitation, create a localized "positron dipole" near the ground. This structure accelerates positrons downward toward the detector while modifying electron transport. The measurements provide evidence that thunderstorms can temporarily create localized regions of atmospheric antimatter acceleration.

## Key Points

• Time-resolved Aragats observations reveal delayed 511-keV positron enhancement relative to ordinary electron/gamma-ray TGE maxima.

• Strong positron enhancement occurs during mature LPCR conditions characterized by positive NSEF, graupel precipitation, and extremely low cloud base.

• A dual-dipole electrodynamic model explains positron acceleration by an LPCR–Earth dipole distinct from the ordinary electron-accelerating thundercloud dipole.

## 1. Introduction

Thunderstorm Ground Enhancements (TGEs) are bursts of secondary cosmic rays observed at the Earth's surface during thunderstorms. Long-term measurements at Aragats showed that thundercloud electric fields can accelerate electrons to relativistic energies and produce substantial enhancements in gamma rays, electrons, and, in some cases, neutrons (Chilingarian et al., 2010; Chilingarian et al., 2011). These observations provided direct experimental support for relativistic runaway electron avalanches (RREAs, Gurevich et al., 1992; Alexeenko et al., 2001; Babich et al., 2002; Dwyer, 2003) developing in the atmosphere and established thunderstorms as natural high-energy particle accelerators.

Subsequent studies have shown that the charge structure of thunderclouds plays a critical role in determining the properties of TGEs and lightning initiation. In addition to the classical Wilsonian dipole (Wilson, 1924), consisting of the main negative layer (MN) and its mirror charge induced at Earth, observations and modeling have demonstrated the importance of the Lower Positive Charge Region (LPCR) near the cloud base (Kuettner, 1950; Williams, 1989; Chilingarian et al., 2012; 2017). The LPCR is closely linked to graupel microphysics and transient thundercloud electrification processes. The emergence of the LPCR substantially modifies the lower-atmosphere electric structure and may strongly affect charged-particle transport near the ground. Measurements performed at Aragats also revealed that thunderstorms significantly enhance atmospheric radioactivity through the radon circulation effect (Chilingarian, Hovsepyan, and Sargsyan, 2020). Radon progeny isotopes attached to atmospheric aerosols are transported upward by strong atmospheric electric fields and turbulent circulation, producing enhancements of characteristic gamma-ray lines, i.e., the 214Pb and 214Bi.

The first evidence of thunderstorm-related positron enhancement was reported in Chilingarian and Sargsyan (2024), linking an increase in the 511-keV annihilation line to LPCR development and a positive near-surface electric field (NSEF). Subsequently, Chilingarian, Sargsyan, and Zazyan (2024) reported an extreme positron event on Aragats, with a large increase in positron-related flux, and proposed the formation of a localized LPCR–Earth dipole that accelerates positrons downward. However, those studies relied mainly on integrated spectral analysis and did not resolve the temporal evolution of positron enhancement relative to that of ordinary electron/gamma-ray TGE. The 16–17 May 2026 events provide the first time-resolved observations demonstrating that positron acceleration evolves dynamically and can become decoupled from the main runaway-electron avalanche.

Independent observations also support the existence of thunderstorm-related positron phenomena in the upper dipole. Airborne ADELE measurements revealed transient positron clouds within thunderstorms, associated with strong annihilation signatures (Briggs et al., 2011). Recently, the ESTHER collaboration reported an enhancement in the 511-keV energy region during thunderstorms at Mount Etna, providing additional evidence that positron-related atmospheric processes may be observable at ground level under high-electric-field conditions (Quartieri et al., 2025).

Despite these advances, several key physical questions remain unresolved. First, previous studies relied mainly on integrated spectra or statistical averages and therefore could not investigate the temporal evolution of positron signatures relative to the development of ordinary electron/gamma-ray TGE.

Second, the relationship between LPCR maturity and positron enhancement remained poorly constrained. Third, it remained unclear how to separate positron-related enhancement from radon progeny enhancement transported by atmospheric aerosols during thunderstorms.

The 16–17 May 2026 events observed at Aragats offer a unique opportunity to address these problems. To interpret these observations, we introduce a dual-dipole electrodynamic model. In this framework, the large-scale electron dipole is formed by the MN and its broad positive mirror charge induced at the Earth's surface, accelerating electrons downward over a large area. Simultaneously, a smaller LPCR–Earth dipole is formed by the LPCR and its localized negative mirror charge directly beneath the LPCR footprint. Because the LPCR is much smaller than the MN in both spatial extent and total charge, positron acceleration occurs only locally beneath the LPCR, whereas electrons accelerated in the upper electron dipole are decelerated when entering the lower positron dipole.

In this paper, we demonstrate that positron enhancement is a dynamically evolving electrodynamic phenomenon directly connected to LPCR maturity, graupel microphysics, and the thundercloud charge geometry. We show that positron TGEs represent a distinct subclass of TGEs characterized by strong positive NSEF, mature LPCR development, delayed 511-keV enhancement, and positron acceleration in the LPCR–Earth positron dipole.

The experimental facilities of the Aragats research station and the TGE selection methodology were described in detail in Chilingarian et al. (2023; 2024a; 2024b; 2026).

## 2. Observations and event context for the 16-17 May 2026 TGEs

### 2.1 Data

Positron TGEs were observed on 16 May 2026 around 09:34 and on 17 May 2026 around 14:26. At 17:03, we registered a classic electron TGE with moderate positron contamination. Continuous observations include NSEF, particle fluxes by STAND3, SEVAN detectors, weather, and lightning-distance records for 16-17 May, along with three integrated energy spectrograms highlighting enhanced spectral lines in the 5-3000 keV range. All three periods show increased particle concentrations under stormy, disturbed conditions. The spectra reveal enhanced 511-keV and radon progeny windows, particularly 214Pb and 214Bi.
Figure 1 provides the broader environmental context for the 16-17 May 2026 thunderstorm sequence. The TGEs occurred during a prolonged thunderstorm episode characterized by NSEF disturbances, very high relative humidity, a low cloud base, and repeated particle flux enhancements in the surface detectors. Positive NSEF at Aragats is interpreted as the surface signature of a mature LPCR near the cloud base. The lower positive charge becomes sufficiently developed to reverse the near-ground electric-field polarity and form a local electrostatic system with its induced mirror charge at the surface. The graupel precipitation reported by station staff supports this interpretation, as graupel is a key hydrometeor in the development of LPCRs.
The TGE overview in Figure 1 presents time series of NSEF, TGE particle fluxes, temperatures, and dew points as a coherent sequence of thunderstorm parameters. On 16 May, we observed a very large positive NSEF with no enhancement in electron/gamma-ray flux. The first TGE on 17 May is characterized by pronounced disturbances in the NSEF, including both positive and negative episodes, with moderate electron/gamma-ray flux. The second TGE on 17 May demonstrates a large electron/gamma-ray flux in the absence of positive NSEF. Thus, these three events cover various conditions of field disturbances and particle flux enhancements.
Cloud-base height was estimated from the surface temperature and dew-point temperature using the standard spread approximation $H_{cb} = 122 \times (T - T_d)$, where T and $T_d$ are in degrees Celsius, and $H_{cb}$ is in meters above the Aragats station level. This value approximates the vertical distance between the detector level and the cloud base. For the three TGE intervals, the very small spreads (0.3-0.4 C) correspond to cloud bases only about 38-50 m above the station, consistent with very low cloud-base and graupel conditions.

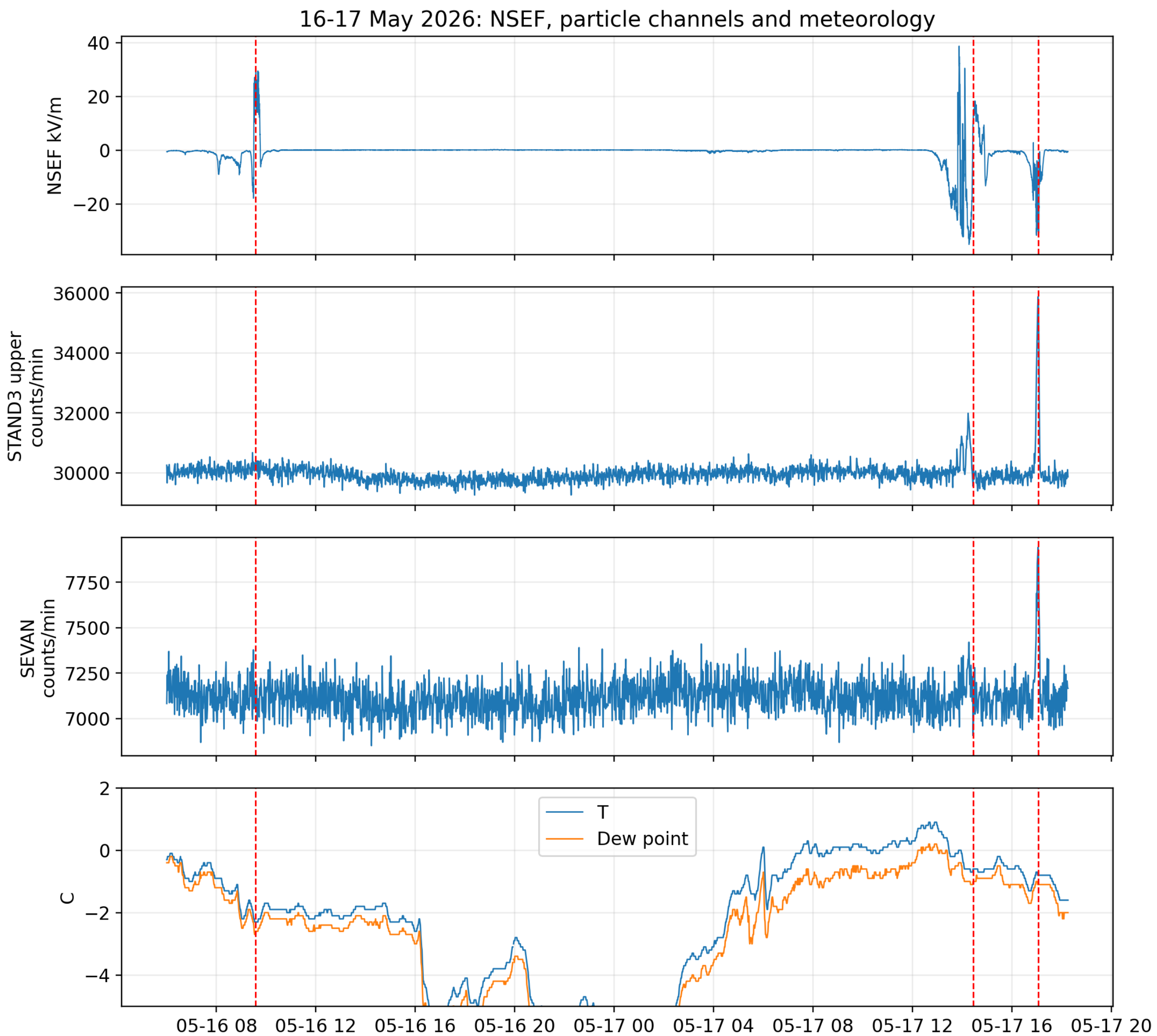

**Figure 1. Overview of meteorological, electric-field, and particle conditions during 16-17 May 2026. NSEF, STAND3, SEVAN, and weather parameters show the thunderstorm environment in which the three TGEs occurred. The red dashed line denotes the positron-flux maximum.**

2.2 Spectral Characteristics of Selected Events

In Figure 2, we show the spectrograms of 3 selected events measured by the precise gamma spectrometer RT 56. Thunderstorm electric fields also lift aerosols carrying radon progeny, producing strong enhancements in the 214Pb and 214Bi lines. Therefore, the 511-keV line must be evaluated together with the 214Pb at 352 keV, the 214Bi at 609 keV, and the broader radon-progeny continuum. The radon progeny are attached to aerosols of both charge signs, and their enhancement is controlled mainly by the field magnitude and atmospheric transport.

In the 16 May spectrum, the 511-keV enhancement is comparable to the radon-progeny enhancement. This agrees with the time-series interpretation: the event was dominated by LPCR/radon-circulation conditions and did not produce a significant electron TGE. The enhancement of the 511-keV region of interest (ROI) is statistically significant in the selected window, but it is not cleanly separated from the radon response.

In the 17 May 14:26 spectrum, the 511-keV window shows the strongest enhancement, about 35%. This is significantly larger than the average radon progeny response. This event is therefore the clearest positron TGE. In the 17 May 17:03 spectrum, the 511-keV enhancement is moderate and occurs simultaneously with a strong ordinary electron TGE.

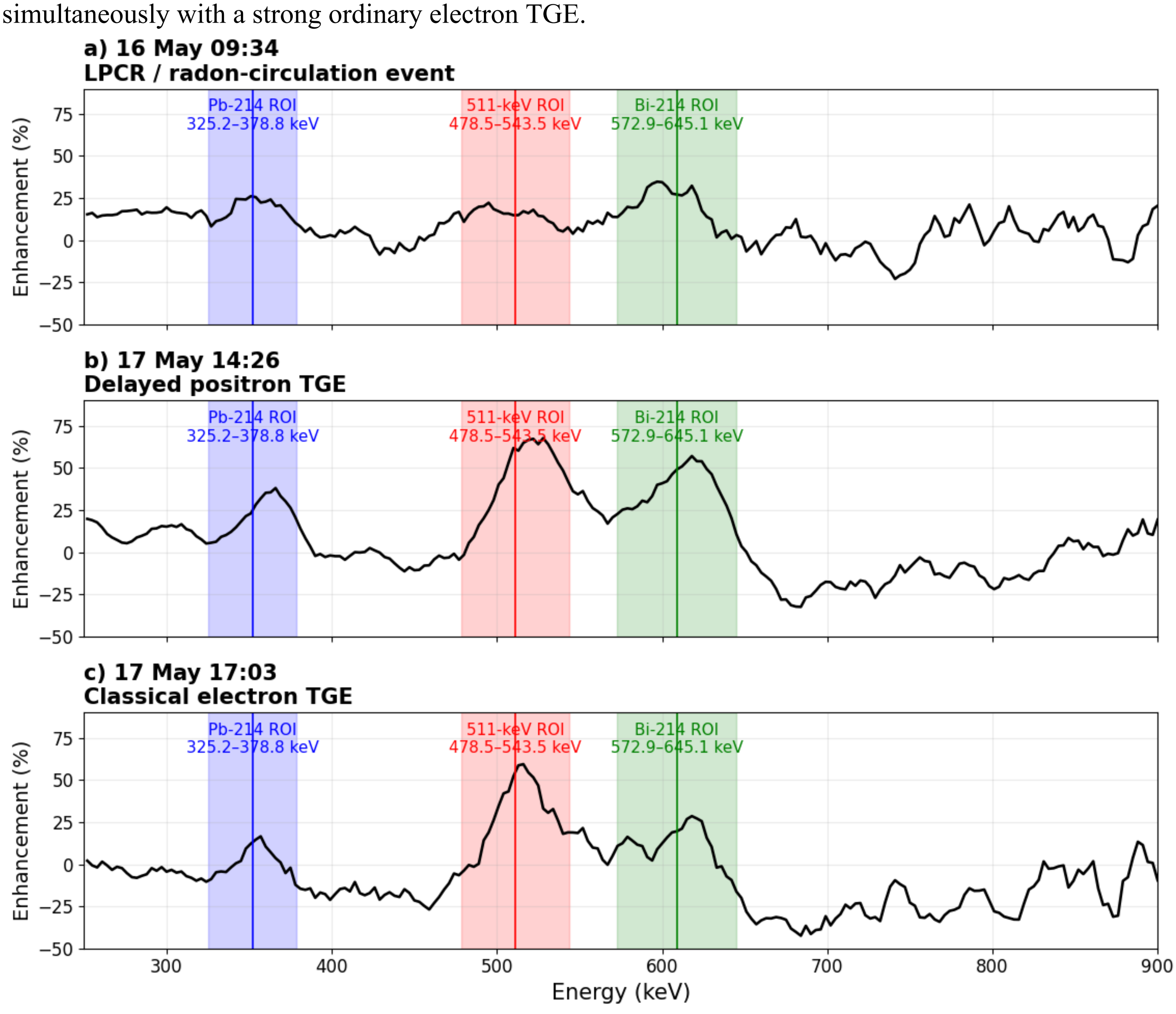

**Figure 2. Gamma-ray spectra measured by the RT 56 spectrometer for the three selected thunderstorm events. Shaded vertical bands indicate the fixed regions of interest (ROIs) used by the RT56 spectrometer for quantitative integration of gamma-ray lines:** Pb-214 (325.2–378.8 keV), positron annihilation (478.5–543.5 keV), and Bi-214 (572.9–645.1 keV)**.**

### 2.3. Meteorological, electric-field, and particle context

Figure 3 compares NSEF and particle fluxes measured by STAND3 and SEVAN detectors in the same time windows around the three TGEs. The red dashed line marks the maximum of the 511-keV window, while the green dotted line marks the electron/gamma-ray channel maximum. This format makes the time shift between the ordinary TGE response and the positron response directly visible. For 16 May, the figure confirms that there was no significant electron TGE. The STAND3 and SEVAN particle curves do not show avalanche-type enhancement.

For the 17 May event, the separation between the particle maximum and the 511-keV maximum is the most significant result. The STAND3 maximum occurs near 14:14, and the SEVAN particle response peaks shortly after, whereas the 511-keV maximum occurs later, at 14:26. This delay indicates that the annihilation-line maximum appears as the NSEF evolves toward a strong positive state, signaling LPCR maturity and a strengthening of the lower positron dipole. The timing, therefore, supports the dual-dipole interpretation: the electron/gamma-ray TGE is controlled mainly by the broad MN-Earth electron dipole, while the later 511-keV response is controlled by the LPCR-Earth positron dipole.
The second event on 17 May (17:03 UT) is a classical electron TGE. The NSEF does not enter the positive NSEF domain. In this event, the STAND3 and SEVAN enhancements were much stronger, and the 511-keV enhancement was moderate. Both enhancements occurred simultaneously. This indicates the simultaneous development of the ordinary electron/gamma-ray TGE and the positron-related process. Thus, the enhanced positron flux is due to the increased TGE gamma-ray flux, which produces additional electron-positron pairs.

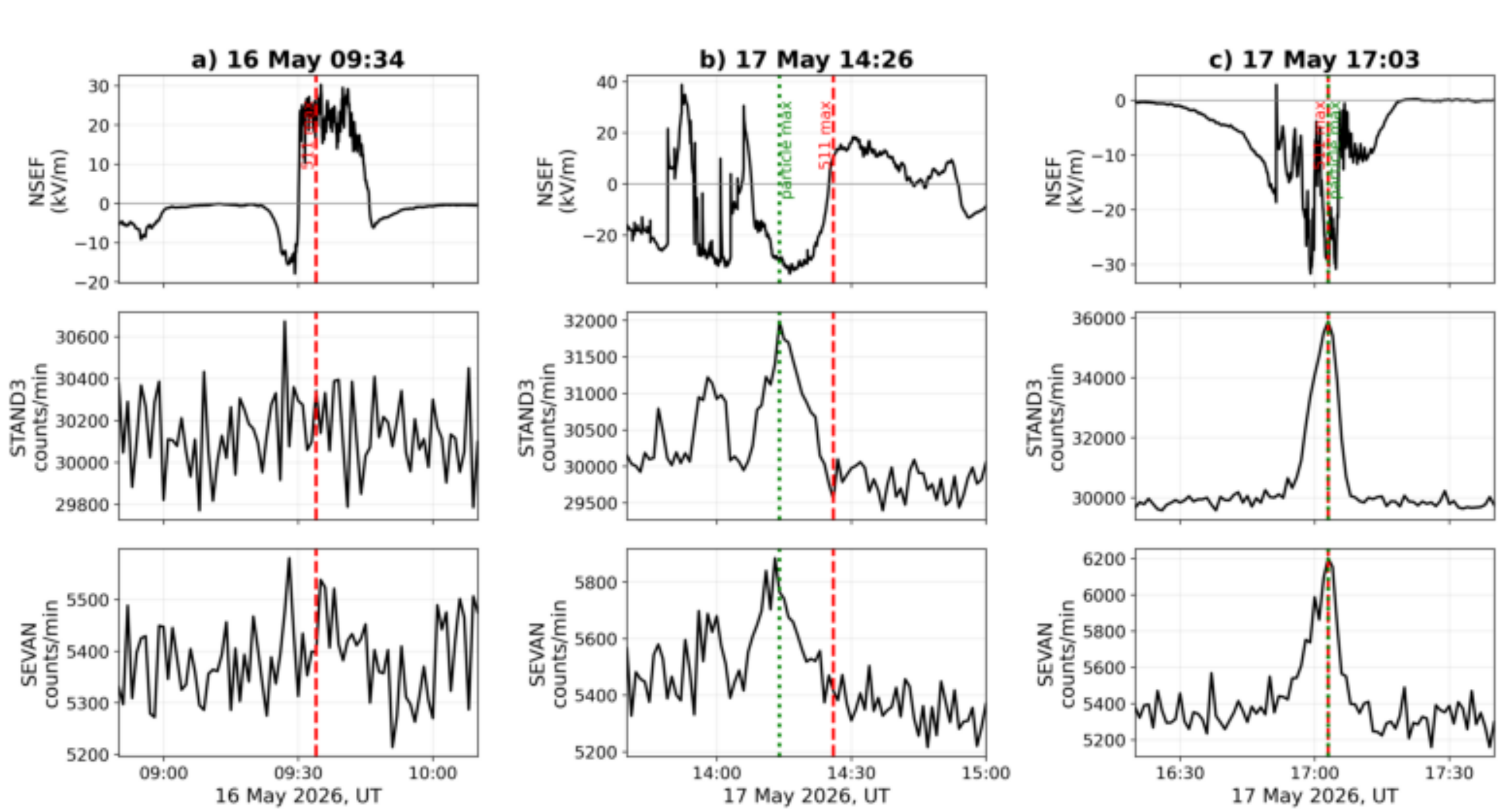

**Figure 3. Temporal evolution of the NSEF, and particle fluxes measured by STAND3 and SEVAN detectors during the three representative thunderstorm events. Red dashed lines indicate the maximum enhancement of the 511-keV annihilation window measured by the RT56 spectrometer, whereas green dotted lines indicate the maximum of the electron TGE measured by STAND3 and SEVAN detectors.**

### 3. Dual-Dipole Electrodynamics of Positron TGEs Above Aragats

The 16–17 May 2026 events reveal dynamical restructuring of the thundercloud electric field through shifts in the timing of particle and radiation enhancements. Table 1 summarizes the meteorological, electric-field, particle-flux, and spectral-line conditions for the three selected events. The resulting cloud-base distances of 35-50 m indicate that the lower charged cloud boundary was extremely close to the detectors. This strongly favors direct coupling between the LPCR, its mirror charge at the ground, and the particle flux measured by SEVAN, RT56, and STAND3. The table shows that the 511-keV annihilation enhancement evolves differently from the ordinary electron/gamma-ray TGE measured by the STAND3 and SEVAN detectors.

The 16 May event represents the weakest ordinary TGE regime. Almost no electron enhancement was detected by STAND3 or SEVAN, while the 511-keV annihilation line increased by about 15%. This result indicates that mature LPCR conditions can produce localized positron enhancement in the absence of a runaway-electron avalanche.

The first 17 May event exhibits the clearest temporal separation between positron and electron/gamma-ray maxima. The ordinary TGE peaked at 14:14, whereas the maximum 511-keV enhancement occurred approximately 12 minutes later, at 14:26. This delay demonstrates that positron acceleration is not controlled by the intensity of the electron-gamma ray runaway avalanche. Instead, the strongest positron enhancement developed during the later stage of lower-cloud electrodynamic evolution associated with LPCR maturity and strong positive NSEF.

**Table 1. Electric-field, cloud-base, particle-flux, and spectral-line context for the three events.**

| Time | Delay of positrons | NSEF kV/m | Cloud-base m | STAND3 enhancement (%) | 511 keV % | Bi214 609 keV % |
|---|---|---|---|---|---|---|
| 16 May 2026 09:34 | - | 30 to 10 | 37 | 0 | 14.82 | 19.01 |
| 17 May 2026 14:26 | 12 | 15 to -19 | 50 | 5.9 | 35.34 | 29.92 |
| 17 May 2026 17:03 | 0 | -2 to -31 | 37 | 20 | 24.05 | 10.97 |

The second 17 May event is a classical electron TGE accompanied by a simultaneous 511-keV enhancement. In this case, the positron and electron/gamma-ray maxima occurred together near 17:03. The NSEF was strongly negative, and there is no evidence of mature LPCR-controlled positron acceleration. Therefore, the observed enhancement of the 511-keV ROI is most naturally explained by the combined contribution of genuine annihilation photons produced by TGE-induced pair production and contamination from the Compton continuum of enhanced Bi-214 gamma radiation.
This event serves as an important contrasting case: unlike the first two events, the positron signal follows the ordinary runaway-electron avalanche rather than evolving independently during positive NSEF.
Figure 4 summarizes the physical model used to interpret the observations. The model contains four charge regions and two physically distinct dipoles. The large-scale electron dipole is formed by the main negative layer and the broad positive mirror charge induced at the Aragats surface. This dipole spans a large area and accelerates electrons downward, producing ordinary electron/gamma-ray TGEs through runaway multiplication and bremsstrahlung.
The second dipole forms between the LPCR near the cloud base and a negative mirror charge directly beneath the LPCR footprint at the surface. The positron dipole is spatially limited; it does not replace the electron dipole across the entire storm area; instead, it locally modifies the lowest part of the field structure under the LPCR. In this region, positrons are accelerated downward, while electrons entering it are decelerated.

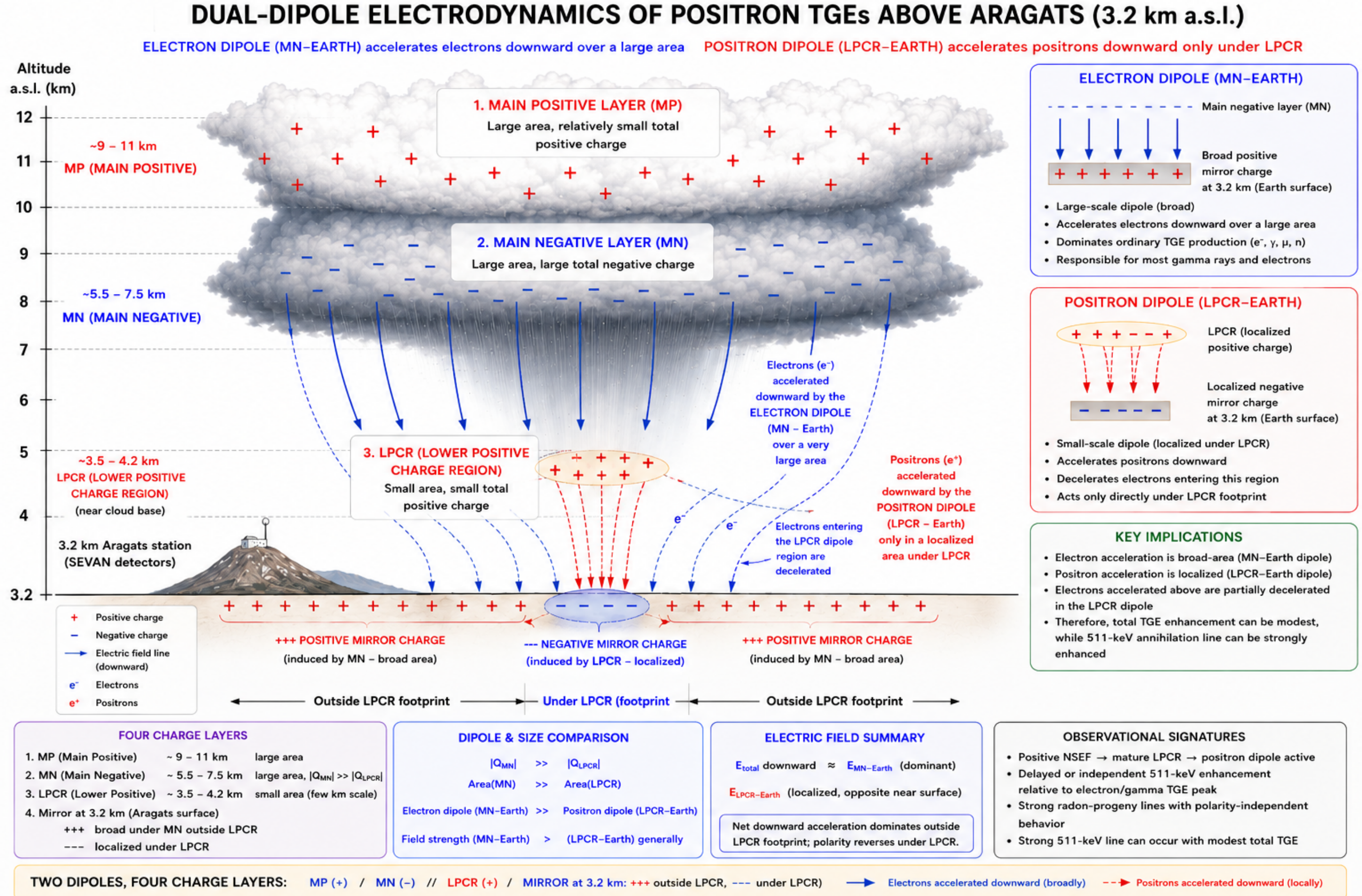

**Figure 4 Dual-dipole electrodynamic model of positron TGEs above Aragats station (3.2 km a.s.l.). The broad MN–Earth electron dipole accelerates electrons downward over a large area, thereby dominating electron TGE production. Simultaneously, the localized LPCR–Earth positron dipole accelerates positrons downward beneath the LPCR footprint, while decelerating electrons entering the same region. This geometry explains the delayed 511-keV enhancement and the coexistence of strong annihilation signatures with a modest total TGE flux.**

This geometry explains the key observational features. A strong 511-keV enhancement can occur when the total electron/gamma-ray enhancement is modest because the positron dipole selectively favors positron transport and annihilation signatures. It also explains why the 511-keV maximum can lag the electron/gamma-ray maximum: the electron TGE responds to the broad MN-Earth accelerator, whereas the 511-keV line responds to the later maturity and geometry of the LPCR-Earth positron dipole. The model therefore unifies the observed positive NSEF, graupel precipitation, low cloud base, timing shifts, and spectral line enhancements into a single electrodynamic scenario. Thus, the measurements distinguish LPCR-driven positron acceleration from secondary positron production within classical electron TGEs.

## 4. Discussion and conclusions

The present observations demonstrate that positron enhancement during thunderstorms arises from several physically distinct mechanisms rather than from a single universal RREA process. The combined analysis of NSEF, meteorological conditions, particle fluxes measured by STAND3, SEVAN detectors, and RT 56 spectrometry reveals that the enhancement of the 511-keV annihilation line depends strongly on the evolving lower-cloud charge structure.

The strongest evidence for independent positron electrodynamics comes from the delayed 511-keV enhancement observed during the first 17 May event. The temporal offset between the maximum electron/gamma-ray flux and the positron maximum shows that positron acceleration is not proportional to the intensity of the runaway-electron avalanche. Instead, the delay indicates that the lower atmospheric charge structure continued evolving after the ordinary TGE peaked. Strong positive NSEF, graupel precipitation, and an extremely low cloud base indicate the formation of a mature LPCR and a localized lower dipole between the LPCR and its mirror charge at the Earth's surface.

Within this lower dipole, positrons are accelerated downward toward the detector, whereas electrons entering the region are decelerated. Consequently, the total electron/gamma-ray enhancement can remain moderate while the annihilation line becomes strongly enhanced. This geometry naturally explains the observed temporal decoupling between the ordinary TGE maximum and the positron maximum. The observations indicate that enhancement of the 511-keV ROI can arise from two distinct physical regimes. During mature LPCR conditions associated with strong positive NSEF, the annihilation line enhancement reflects direct positron acceleration within the localized LPCR–Earth dipole. In contrast, during strong negative NSEF, simultaneous enhancement of the electron avalanche and the 511-keV line is more naturally explained by additional pair production driven by the intense gamma-ray flux of the ordinary TGE itself. Spectrometric measurements further show that enhancements of radon progeny and the annihilation line depend differently on thunderstorm polarity. Previous Aragats observations showed that negative aerosols lift radon progeny more efficiently than positive ones (Chilingarian, Sargsyan, and Zazyan, 2024). Therefore, radon-progeny enhancement depends mainly on aerosol transport, precipitation, turbulence, and electric-field-driven circulation in the lower atmosphere. In contrast, the 511-keV annihilation line depends much more strongly on the direction of charged-particle acceleration and thundercloud charge geometry. This distinction explains why radon progeny and positron enhancements do not necessarily evolve synchronously.

The present results support a dual-dipole electrodynamic interpretation of thunderstorm particle acceleration. The large-scale MN–Earth dipole accelerates electrons downward over a broad atmospheric region and dominates ordinary TGE production. Simultaneously, a localized LPCR–Earth dipole may accelerate positrons downward only beneath the LPCR footprint. Thus, the 2024 Aragats positron studies established the existence and magnitude of positron enhancement during thunderstorms, whereas the 2026 observations reveal its time-dependent electrodynamic origin. The delayed positron maximum observed on 17 May demonstrates that localized positron acceleration can become partially decoupled from the main runaway-electron avalanche.

The present measurements, therefore, extend the standard TGE paradigm from a single electron-dominated accelerator to a dynamically evolving multi-layer atmospheric accelerator system in which localized positron acceleration may arise during mature LPCR conditions.

**Data Availability Statement**

The full set of TGEs is available in both numerical and graphical formats at the cosmic ray division's database of the Yerevan Physics Institute: http://crd.yerphi.am/adei/. The WIKI section of the database contains instructions on selecting appropriate detectors and times, as well as on uploading data.